\documentclass[lettersize,journal]{IEEEtran}
\normalsize

\makeatletter
\def\ps@IEEEtitlepagestyle{%
  \def\@oddhead{%
    \vbox{%
      \hbox to \textwidth{\parbox[t]{\textwidth}{%
        This work has been submitted to the IEEE for possible publication. Copyright may be transferred without notice,
        after which this version may no longer be accessible.%
      }}%
      \vspace{2pt}%
      \hrule height 0.4pt%
    }%
  }%
  \def\@evenhead{\@oddhead}%
  \def\@oddfoot{}%
  \def\@evenfoot{}%
}
\makeatother

\usepackage{amsmath,graphicx}
\usepackage{url}
\usepackage{bm}
\usepackage{amsfonts}
\usepackage{amssymb}
\usepackage{amsthm}
\usepackage{dsfont}
\usepackage{algorithm,algorithmic}
\usepackage{caption,subcaption}
 \usepackage[table,xcdraw]{xcolor}
\usepackage[normalem]{ulem}

\usepackage{tikz}

\usepackage{tipa}

\usepackage{tikz}
\usetikzlibrary{arrows}
 \usepackage[noadjust]{cite}
 
\usepackage{cite,multirow}
\usepackage{tcolorbox}
\usepackage[labelformat=simple]{subcaption}
\usepackage{algorithm,algorithmic}

\begin{document}

\title{MJSAC: McCormick Relaxation-based Waveform Design for Joint Sensing and Communication}
\author{Bodhibrata Mukhopadhyay,~\IEEEmembership{Member,~IEEE}, Sajid Ahmed,~\IEEEmembership{Senior Member,~IEEE}, Mohamed-Slim Alouini,~\IEEEmembership{Fellow,~IEEE}
\thanks{The first two authors contributed equally to this work. Bodhibrata Mukhopadhyay was with CEMSE, King Abdullah University of Science and Technology (KAUST), Thuwal, 23955-6900, Kingdom of Saudi Arabia. He is now with the department of Electronics and Communication, Indian Institute of Technology Roorkee, Uttarakhand, 247667, India  S. Ahmed, and M.S. Alouini are with the CEMSE, , Thuwal, 23955-6900, Kingdom of Saudi Arabia (email: bodhibrata@ece.iitr.ac.in, sajid.ahmed@kaust.edu.sa, slim.alouini@kaust.edu.sa).}}
\maketitle

\begin{abstract}
In the upcoming 5G Advanced and 6G technologies, joint sensing and communication (JSAC) will play a pivotal role in enabling the simultaneous utilization of hardware and spectrum resources for communication and sensing tasks. While current algorithms primarily focus on designing beampattern invariant covariance matrices for transmitting various symbols for communication, they often overlook the distances among these symbols. While these covariance matrices effectively facilitate ranging operations, they have adverse effects on communication performance. Designing beampattern invariance covariance matrices with maximal distances among themselves poses a challenging non-convex problem. In this paper, we introduce a novel waveform design method based on McCormick relaxation called McCormick-based JSAC (MJSAC). MJSAC sequentially solves an optimization problem to generate a set of covariance matrices by maximizing the distances (Frobenius norm) among themselves while ensuring a consistent beam pattern. Also, MJSAC eliminates the requirement for channel information to generate the covariance matrices. Through simulations, we demonstrate that MJSAC outperforms conventional algorithms, even those utilizing channel information at the transmitter.
\end{abstract}

\section{Introduction}
In the rapidly evolving landscape of wireless communication technologies, joint sensing and communication (JSAC) has become a key focus for researchers and industry professionals~\cite{Chen_JSAC,Terahertz_6G,ISAC_MIMO,Liu_JCR_2020,wei_integrated_2023}. This convergence enables advanced applications in various domains, such as connected vehicles, smart cities, industrial automation, and healthcare \cite{Zhe_joint_pilot_opt}. Joint sensing and communication systems aim to seamlessly integrate radar and communication capabilities within the same system~\cite{Eldar_Joint_Beamforming}. This framework offers numerous advantages, including the same spectrum and hardware resources for communication and sensing, improved efficiency, reduced latency, and enhanced reliability~\cite{Poor_Federated_Ambient_Intelligence}.

The integration of radar and communication can be broadly categorized into three main groups: {1) \textit{Cohabitation}:} In this approach, both radar and communication systems operate within the same frequency band simultaneously. However, this simultaneous operation leads to interference between the radar and communication signals \cite{tian2019mutual}, {2) \textit{Cooperation}:} This category utilizes an opportunistic spectrum access strategy by designating one user as the primary user and another as the secondary user. The secondary user is permitted to access the spectrum only when the primary user is not utilizing it, as outlined in~\cite{Liu_JCR_2020}, and 3) \textit{Codesign}: This scheme is also referred to as JSAC, where both systems are unified and integrated waveforms are designed. The JSAC system efficiently uses spectrum and hardware resources for both sensing and communication tasks at the same time. This paper centers on codesign methodologies, as discussed in previous works~\cite{zubair_generalized_2022,Chiriyath_Radar_comm_convergence,Sana_2023,Liu_Fomni}.

Zhou \textit{et al.}~\cite{ISAC_waveform_Survey} provides a comprehensive review of waveform design approaches for JSAC, categorizing them into communication-centric, sensing-centric, and joint waveform optimization methods. Their work highlights the importance of joint waveform optimization. Zhang \textit{et al.}~\cite{Ch_train_intrigrated_sensing} introduce a novel random pattern-based JSAC scheme that exploits tensor decomposition techniques for target sensing while performing efficient channel estimation in terahertz massive MIMO (mMIMO) systems. This work demonstrates the benefits of structured JSAC waveform patterns in improving system performance while reducing training overhead.



Within the JSAC framework, Fan Liu~\textit{ et al.} proposed a waveform design technique to achieve specific beampattern and transmit data symbols to multiple users~\cite{Liu_Fomni}. Their approach requires the knowledge of the channel information of each user at the transmitter.  Moreover, it involves a computationally intensive operation to generate the waveforms at the transmitter for each data transmission session.
%
%
The covariance matrices are designed to prioritize sensing over communication, with a focus on the area of interest (AOI) for effective target detection. In scenarios like surveillance, concentrating transmission power towards the AOI conserves energy and enhances detection range. Sana~\textit{et al.} illustrated that the beampattern of a covariance matrix depends on the sum of its diagonals rather than its individual elements~\cite{Sana_2023}. Leveraging this, they devised a method to generate an infinite set of covariance matrices, each resulting in an identical beampattern. Depending on the modulation schemes, different symbols can be assigned to different covariance matrices within the set. This innovative approach enables the transmission of diverse covariance matrices (symbols) with identical beampatterns, eliminating the need for channel state information of multiple users at the transmitter. Nevertheless, it has been noted that the communication performance of the system is significantly poor, attributed to the receiver's inability to distinguish the covariance matrices (which are used as symbols). This is because the distance between the covariance matrices is not adequate.


In this work, we address the limitations of the previous research by eliminating the necessity of channel state information at the transmitter and also improving symbol error rate performance by increasing the separation among the covariance matrices. We introduce MJSAC, a novel waveform design algorithm that generates $L$ covariance matrices for $L$-alphabet modulation scheme by maximizing the distances (Frobenius norm) among themselves. Each of the symbols exhibits the same beam pattern. MJSAC tackles the optimization challenge of maximizing a convex function over a spectrahedron. To address this, the problem is reformulated into standard semidefinite programming by linearizing the objective function using the McCormick relaxation.

    %



\section{Signal Model}
\label{Sec_signal_Model}
%
Consider a JSAC system with $N$ transmit antennas communicating with a single-antenna cellular user. If $x_n(t)$ is the transmitted waveform of duration $T$ from the $n^{\text{th}}$ antenna and $h_n(t)$ is the channel gain between the user and the $n^{\text{th}}$ antenna, the demodulated receive samples in discrete form can be written as
\begin{IEEEeqnarray}{c}
\label{Eq:Signal_model}
    \mathbf{y} = \mathbf{X}\mathbf{h} + \bm{\eta},
\end{IEEEeqnarray}
where $\mathbf{h}=h_1\mathbf{a}(\theta) \in \mathds{C}^{N \times 1}$ is the channel vector while $\mathbf{a}(\theta) = \begin{bmatrix}1,e^{j\pi sin(\theta)},\cdots,e^{j (N-1)\pi sin(\theta)} \end{bmatrix}^T$ is the transmit steering vector. The transmitted signal matrix is represented as $\mathbf{X}=[\mathbf{x}_1,\mathbf{x}_2,\dots,\mathbf{x}_N]$, while $\mathbf{x}_i \in \mathds{C}^{M\times 1}$ is the transmitted signal vector from the $i$th antenna and it contains $M$ samples. For simplicity, we omit the explicit mention of the sampling time $T_s$ and refer to sample indices directly. The transmit power of the entire system is considered to be unity, i.e., $\lVert \mathbf{X}\rVert_F^2=1$ $\left(\lVert\cdot\rVert_F \right.$ represents the Frobenius norm of a matrix$\left. \right)$.
The channel is assumed to follow  Rayleigh fading, and it does not change during one symbol transmission. ${\mathbf{y}}=[y(0),y(1),\cdots,y(M-1)]^T \in \mathds{C}^{M \times 1}$ is the received sample vector at the receiver. The noise vector is represented by $\bm{\eta} = \left[ \eta_1,\eta_2,\dots,\eta_M\right] \in \mathds{C}^{M \times 1}$  where $\eta_i \sim C\mathcal{N}(0,\sigma^2)$.

The covariance matrix of the transmitted waveforms is  ${\mathbf{R}} = \mathbf{X}^H\mathbf{X}$, where $\mathbf{R} \in {\mathds{C}}^{N\times N}$. The corresponding waveforms can be generated using $\mathbf{X}=\bm{\Phi}\mathbf{R}^{1/2}$, where $\bm{\Phi}\left(\in \mathds{R}^{M\times N}\right)$ is an orthogonal matrix $\left(\bm{\Phi}^T\bm{\Phi}=\mathbf{I}_N\right)$.
%
%
Now, the received signal in \eqref{Eq:Signal_model} can be written as
\begin{IEEEeqnarray}{c}
\label{Eq:Signal_model_2}
\mathbf{y} = \bm{\Phi} \mathbf{R}^{1/2}\mathbf{h} + \bm{\eta}. 
\end{IEEEeqnarray}
%
%
Within the framework of MIMO radar, it is feasible to generate covariance matrices that yield waveforms directed toward a designated region for target detection. In~\cite{Sana_2023}, authors demonstrated that multiple covariance matrices having the same beampattern can be generated. The covariance matrices yield identical beampattern when the sum of the diagonal and subdiagonal(s) elements of the matrices remains constant and the matrices are positive semidefinite (PSD). This opens up the opportunity to utilize various waveforms not only for sensing but also for communication purposes. In this study, we use the square roots of the covariance matrices as symbols and treat them as points in the constellation diagram $\left( \mathbf{S}_i = \mathbf{R}^{1/2}_i\right)$. The waveform corresponding to the $i^{\text{th}}$ symbol $\mathbf{S}_i$ is expressed as $\bm{\Phi}\mathbf{S}_i$ and the covariance matrix of the transmitted waveform will be $\left(\bm{\Phi}\mathbf{S}_i\right)^T\bm{\Phi}\mathbf{S}_i=\mathbf{R}, \forall i$.
%
%
Here, we consider $L$ alphabet modulation scheme, i.e., the total number of symbol is $L$. So, we need to generate $L$ covariance matrices yielding the same beampattern $\mathbf{R}$. The generation method of the covariance matrices is discussed in Section~\ref{Sec:MJSAC}. 
Assuming the knowledge of the channel coefficients and all covariance matrices (symbols) at the receiver, the transmitted symbols can be estimated by a user using: 
\begin{IEEEeqnarray}{c}
\label{Eq:detector}
\mathcal{\hat{S}} \, =  \, \underset{ \mathbf{S}_i,\atop i=1,\cdots,L}{\min} \, \lVert \mathbf{y} - \bm{\Phi} \mathbf{S}_i\mathbf{h}\rVert^2_F,
\end{IEEEeqnarray}
where $\mathcal{\hat{S}}$ is the detected symbol.
\color{black}
To achieve better detection performance and lower symbol error rate (SER), the symbol matrices must be selected in such a way that they are well separated in $\mathds{R}^{N\times N}$.

%

\section{Covariance Matrix Design: MJSAC}
\label{Sec:MJSAC}

%
%
%
In this study, we assume that the beampattern associated with $\mathbf{R}$ is symmetric, implying that  $\mathbf{R} \in \mathds{R}^{N \times N}$. To facilitate the transmission of $L$ distinct symbols using varying covariance matrices, we generate $L-1$ PSD matrices, denoted as $\mathbf{Z}_i (\in \mathds{R}^{N \times N}), i=1,\dots,L-1$. We assume the $L^{\text{th}}$ symbol to be equivalent to $\mathbf{R}$. The set of  $\mathbf{Z}_i$ are obtained by solving:
\begin{IEEEeqnarray}{lcl}
\label{Eq:Original_opt_plb}
\IEEEyesnumber \IEEEyessubnumber\label{Eq:Original_obj}
 \underset{\mathbf{Z}_i,\atop i=1,\dots,L-1}{\max} && \sum_{i=1}^{L-1} \lVert \mathbf{R} - \mathbf{Z}_i  \rVert^2_F + \sum_{\substack{i,j=1 \\ i\neq  j,i <j}}^{L-1} \lVert \mathbf{Z}_i - \mathbf{Z}_j  \rVert^2_F \label{Eq:Original_objective} \\
\text{subject to} 
& \ \  &  \bigcup_{i=1}^{L-1} \left[\begin{array}{l} 
\text{tr}\left(\mathbf{D}_m\mathbf{Z}_i\right) = \text{tr} \left(\mathbf{D}_m\mathbf{R}\right), \, \\ \hspace{2.85cm} m=1,\dots,N \\
\mathbf{Z}_i(p,p)= \mathbf{R}(p,p), \,   p=1,\dots,N 
\end{array}\right] \IEEEeqnarraynumspace  \IEEEyessubnumber  \label{Eq:Original_Const1}  \\
&& \mathbf{Z}_i \succcurlyeq  \mathbf{0}_{N},   \quad \ i= 1,\cdots,L-1 \IEEEeqnarraynumspace \IEEEyessubnumber \label{Eq:Original_Const2}
\end{IEEEeqnarray}
%
%
where $\text{tr}(\cdot)$ denotes the trace of a matrix and $\mathbf{R} \in \mathbb{R}^{N \times N}$ is the covariance matrix representing the beampattern (omni or directional) associated with the Radar operation. $\mathbf{D}_m$ is a lower triangular matrix where the entries along the $m^{\text{th}}$ diagonal are assigned a value of 1, while the remaining elements are set to 0.
Matrix $\mathbf{D}_m$ is defined as
%
%
\begin{IEEEeqnarray}{c}
   \mathbf{D}_m(i,j) = \left\lbrace\begin{array}{ll} 1 & i = m,m+1,\ldots,N \nonumber \\ &  j=1,2,\ldots,N-m+1 
   \\
    0 & \mathrm{otherwise }.\end{array} \right.
\end{IEEEeqnarray}
 
In problem~\eqref{Eq:Original_opt_plb}, the objective is to construct a set of matrices \( \mathbf{Z}_i \) that are both well-separated from each other and sufficiently distant from the reference covariance matrix \( \mathbf{R} \), within the space \( \mathds{R}^{N \times N} \). The first term in~\eqref{Eq:Original_objective} encourages each \( \mathbf{Z}_i \) to deviate significantly from \( \mathbf{R} \). The second term in~\eqref{Eq:Original_objective} ensures maximum pairwise separation among the \( \mathbf{Z}_i \) matrices themselves. Since the matrices \( \mathbf{Z}_i^{1/2} \) serve as the constellation points in our communication framework, maximizing their mutual distances directly enhances the overall \textit{communication performance}. Constraint~\eqref{Eq:Original_Const1} ensures that the 
beampattern of the waveforms obtained using $\mathbf{Z}_i$ is identical to the beampattern obtained using the given covariance matrix  $\mathbf{R}$~\cite{Sana_2023}. This condition specifically addresses the system's sensing functionality. Also,~\eqref{Eq:Original_Const2} assures that the generated $\mathbf{Z}_i$ are PSD. 

The optimization problem~\eqref{Eq:Original_opt_plb} is neither convex nor concave and can not be solved using off-the-shelf techniques. Also, the number and size of the decision variables and constraints increase exponentially with $L$. Therefore, we solve~\eqref{Eq:Original_opt_plb} in a sequential manner i.e. we first estimate $\mathbf{Z}_1$ such that it maximizes $\lVert\mathbf{R} - \mathbf{Z}_1\rVert^2_F$ and then use the estimated $\mathbf{Z}_1$ to obtain $\mathbf{Z}_2$ such that $\lVert\mathbf{R} - \mathbf{Z}_2\rVert^2_F$ + $\lVert\mathbf{Z}_1 - \mathbf{Z}_2\rVert^2_F$ is maximized. We iterate the process $L$ times to generate $\mathbf{Z}_i$s. The optimization problem to estimate $\mathbf{Z}_k$ assuming  $\mathbf{Z}_1, \dots \mathbf{Z}_{k-1}$ are already estimated is represented as:
\begin{IEEEeqnarray}{lcl}
\label{Eq:One_Term_opt_plb}
\IEEEyesnumber \IEEEyessubnumber\label{Eq:One_Term_obj}
 \underset{\mathbf{Z}_k}{\max} & \ & \lVert \mathbf{R} - \mathbf{Z}_k  \rVert^2_F + \sum_{i=1}^{k-1}\lVert \mathbf{Z}_{i} - \mathbf{Z}_k  \rVert^2_F  \\
\text{subject to} & \ \  & \text{tr}\left(\mathbf{D}_m\mathbf{Z}_k\right) = \text{tr} \left(\mathbf{D}_m\mathbf{R}\right), \,   m=1,\dots,N\IEEEyessubnumber \label{Eq:One_Term_Const1} \IEEEeqnarraynumspace\\ 
\IEEEyessubnumber \label{Eq:One_Term_Const2}
& \ \ & 
\mathbf{Z}_{k}(p,p) = \mathbf{R}(p,p), \, p=1,\dots,N  \\ 
\IEEEyessubnumber  \label{Eq:One_Term_Const3}
 & \ \ & \mathbf{Z}_k \succcurlyeq  \mathbf{0}_{N} 
\end{IEEEeqnarray}
The square of the Frobenius norm of a matrix can be expressed as the sum of the squares of its elements. Therefore~\eqref{Eq:One_Term_opt_plb} can be written as:
\begin{IEEEeqnarray}{lcl}
\label{Eq:One_Term_opt_plb_V2}
\IEEEyesnumber \IEEEyessubnumber
 \underset{\mathbf{Z}_k}{\max} & \ &  \sum_{p,q=1}^{N} \mathbf{Z}_k(p,q)^2 - \nonumber \\ & \ &  \frac{2}{k}\sum_{p,q=1}^{N} \left(\mathbf{R}(p,q)+\sum_{i=1}^{k-1} \mathbf{Z}_i(p,q)  \right)  \mathbf{Z}_k(p,q) \label{Eq:One_Term_opt_plb_V2_obj} \IEEEeqnarraynumspace\\ \IEEEyessubnumber \text{subject to} & \  & ~\eqref{Eq:One_Term_Const1},\eqref{Eq:One_Term_Const2},\eqref{Eq:One_Term_Const3}     \nonumber
\end{IEEEeqnarray}
The optimization problem in \eqref{Eq:One_Term_opt_plb_V2} poses a significant challenge, as it is inherently non-tractable and cannot be addressed through conventional optimization methods due to its focus on maximizing a convex function ($\lVert \cdot \rVert^2_F$)~\cite{selvi_convex_2022}. Solving this optimization is not trivial since convex functions generally do not have a global maximum. Despite the constraint set being a spectrahedron, the problem remains challenging due to its non-linear objective function (refer~\eqref{Eq:One_Term_opt_plb_V2_obj}). Therefore, to transform the problem into an SDP~\cite{boyd2004convex,dimitri1999nonlinear}, we use McCormick relaxation~\cite{scott2011mccormick,mitsos2009mccormick} to linearize the objective function. McCormick relaxation involves introducing bounded auxiliary variables to represent the decision variable responsible for the non-linearity of the function. By using this technique we approximate the decision variable $\mathbf{Z}_k(p,q)^2, \forall p,q$ in~\eqref{Eq:One_Term_opt_plb_V2_obj} with an auxiliary variable $\mathbf{w}(i), i=1,\cdots,N^2 \in \mathds{R}$. By doing so, the originally non-convex problem~\eqref{Eq:One_Term_opt_plb_V2} can be reformulated into a convex optimization problem, making it more amenable to efficient solution methods. Also each pair of $\left\{\mathbf{w}(i),\mathbf{Z}_k(p,q)\right\}$ is bounded in $\mathbf{R}^2$ by three inequalities (refer~\eqref{Eq:relax_oneterm_obj_const}). Thus using  McCormick relaxation  we convert~\eqref{Eq:One_Term_opt_plb_V2} into a convex optimization problem:
\begin{IEEEeqnarray}{ll}
\label{Eq:relax_McCormick} 
\IEEEyesnumber \IEEEyessubnumber
\label{Eq:relax_oneterm_obj}    
 & \underset{\mathbf{Z}_k,{\bf w}}{\max}  \,\,\,\, \sum_{i=1}^{N^2} \mathbf{w}(i) - \nonumber \\ &  \hspace{1cm} \frac{2}{k}\sum_{p,q=1}^{N} \left(\mathbf{R}(p,q)+\sum_{i=1}^{k-1} \mathbf{Z}_i(p,q)  \right)  \mathbf{Z}_k(p,q) \\ \IEEEyessubnumber 
 & \text{subject to}  ~\eqref{Eq:One_Term_Const1},\eqref{Eq:One_Term_Const2},\eqref{Eq:One_Term_Const3}  \nonumber \\
& \,\,  \bigcup_{i=1}^{N^2} \left[\begin{array}{l} 
\mathbf{w}(i) \geq 2\alpha \mathbf{Z}_k(p,q)-\alpha^2 \\
\mathbf{w}(i) \geq 2\beta \mathbf{Z}_k(p,q)-\beta^2 \\
  \mathbf{w}(i) \leq \alpha \mathbf{Z}_k(p,q)+\beta \mathbf{Z}_k(p,q)-\alpha\beta 
  \end{array}\right] \forall \{p,q\}  \IEEEeqnarraynumspace  \label{Eq:relax_oneterm_obj_const} \\ \IEEEyessubnumber
& \,\, \alpha \leq \mathbf{w}(i) \leq \beta, \, i=1,\dots,N^2 \label{Eq:relax_oneterm_obj_const_2}
\end{IEEEeqnarray}
The optimization problem in~\eqref{Eq:relax_McCormick} has a linear objective function and can be solved using standard SDP solvers. To solve~\eqref{Eq:relax_McCormick}, the constraint~\eqref{Eq:relax_oneterm_obj_const_2} can be omitted since the bounds on $\mathbf{w}(i)$ are already accounted for by~\eqref{Eq:relax_oneterm_obj_const}. The steps to generate covariance matrices using~\eqref{Eq:relax_McCormick} are given in Algorithm~\ref{Algo:MJSAC}.

\begin{figure}[b]
\centering
\includegraphics[width=.9\linewidth,keepaspectratio]{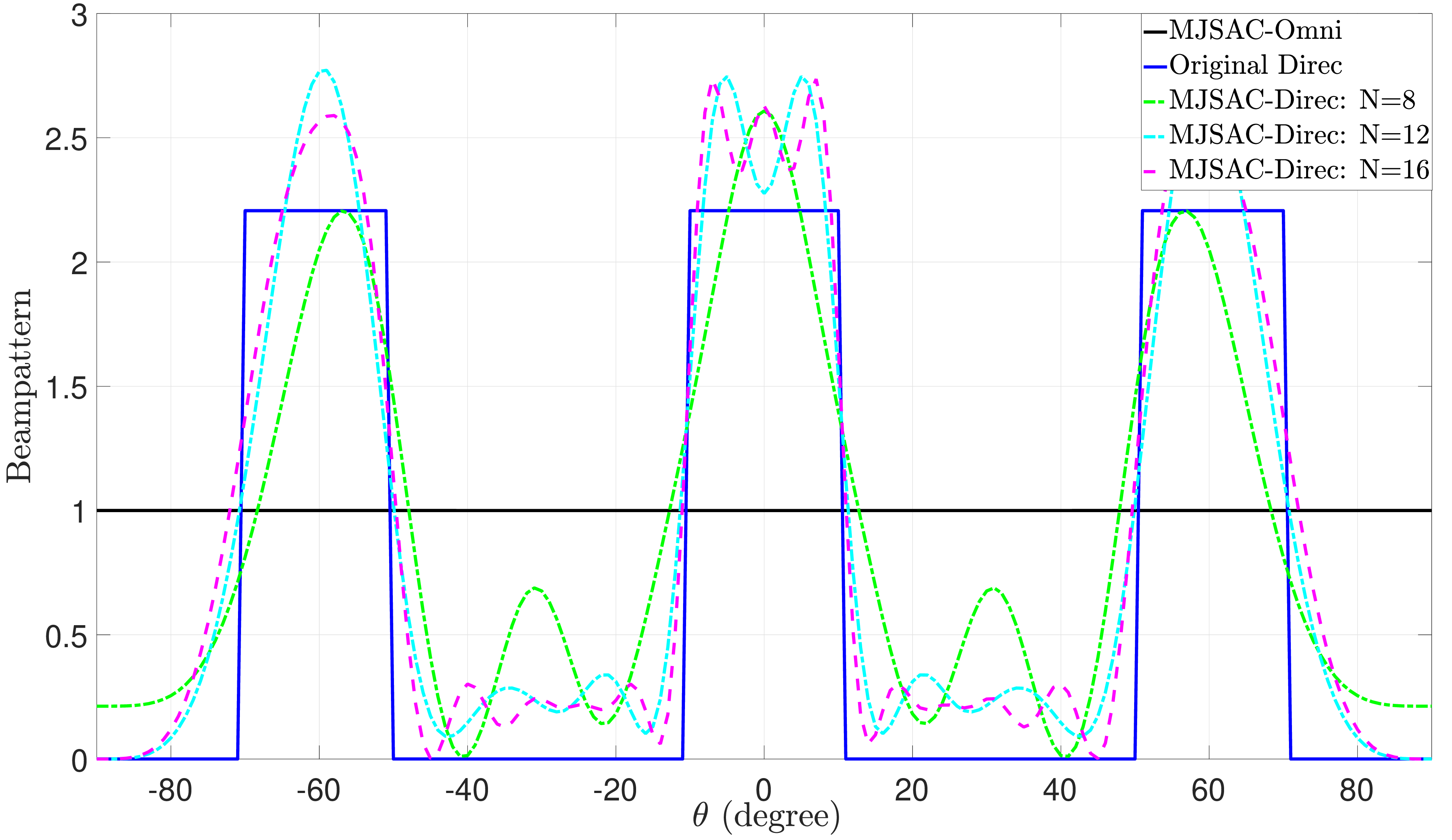}
\caption{Beampattern obtained by MJSAC for different $N$.}
\label{fig:Beam_pattern}
\end{figure}

\begin{algorithm}[!t]
\caption{MJSAC: Generate $L$ Symbols}
\label{Algo:MJSAC}
\begin{algorithmic}[1]
\renewcommand{\algorithmicrequire}{\textbf{Input:}}
\renewcommand{\algorithmicensure}{\textbf{Output:}}
\REQUIRE $\mathbf{R} \in 
\mathbb{R}^{N \times N}$
\ENSURE $\mathbf{S}_i \in 
\mathbb{R}^{N \times N}$, \, $i=1,\dots,L$ 
\FOR{$\text{k}=1:L-1$}
    \STATE Compute $\mathbf{Z}_k$ using~\eqref{Eq:relax_McCormick} 
    \STATE $\mathbf{S}_k       = \mathbf{Z}_k^{1/2}$
\ENDFOR
\STATE $\mathbf{S}_L   = \mathbf{R}$
\RETURN $\mathbf{S}_i, \, i=1,\dots,L$
\end{algorithmic}
\end{algorithm}

\subsection{{Convergence of Algorithm~\ref{Algo:MJSAC}}}
\label{sec_convergece}
Algorithm~\ref{Algo:MJSAC} employs a sequential approach to solve the convexified version of the optimization problem~\eqref{Eq:Original_opt_plb}  using McCormick relaxation and SDP technique. Each subproblem within Algorithm~\ref{Algo:MJSAC}, represented by optimization problem~\eqref{Eq:relax_McCormick}, is convex due to the linearization introduced by McCormick relaxation. This transformation converts the original non-convex problem into an SDP with a \textit{linear objective function} and convex constraints. The feasibility of the optimization is further supported by the bounded nature of the solution space (refer~\eqref{Eq:One_Term_Const1},\eqref{Eq:One_Term_Const2},\eqref{Eq:One_Term_Const3}). This bounded feasible region ensures that the sequence of \( \mathbf{Z}_k \) matrices does not diverge. Additionally, from a theoretical standpoint, convex optimization literature confirms that SDP problems solved using interior-point methods are guaranteed to converge to a global optimum under standard convex conditions~\cite{boyd2004convex,dimitri1999nonlinear}. Since each iteration in Algorithm~\ref{Algo:MJSAC} solves an SDP problem with well-defined convex constraints, convergence is assured for each subproblem. Moreover, the transformation from~\eqref{Eq:One_Term_opt_plb_V2_obj} to~\eqref{Eq:relax_oneterm_obj} via McCormick relaxation results in a tight approximation at the corners or boundaries of the feasible domain. This is because McCormick relaxation provides the best possible convex underestimator and/or concave overestimator over a bounded region~\cite{scott2011mccormick, mitsos2009mccormick}. In~\eqref{Eq:relax_McCormick}, since the objective function is linear, the optimal solution is attained at the boundary defined by the constraints of~\eqref{Eq:relax_McCormick}.

\begin{figure*}[h]
\begin{subfigure}[]{.33\textwidth}
\centering
\includegraphics[width=1.8\linewidth,keepaspectratio]{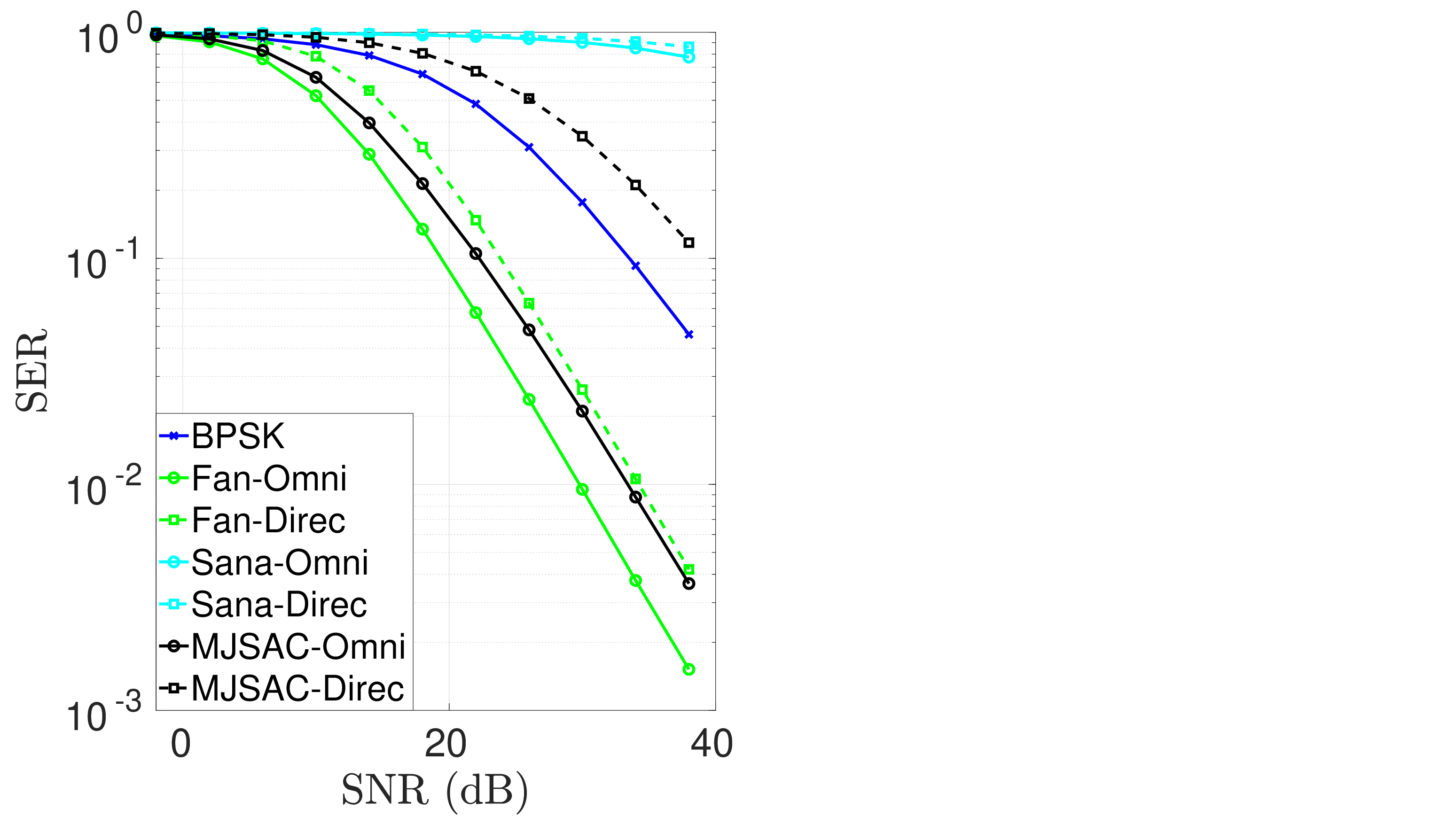}
\caption{Number of antenna is 8.}
\label{fig:no_antennae8}
\end{subfigure}
\begin{subfigure}[]{.33\textwidth}
\centering
\includegraphics[width=1.8\linewidth,keepaspectratio]{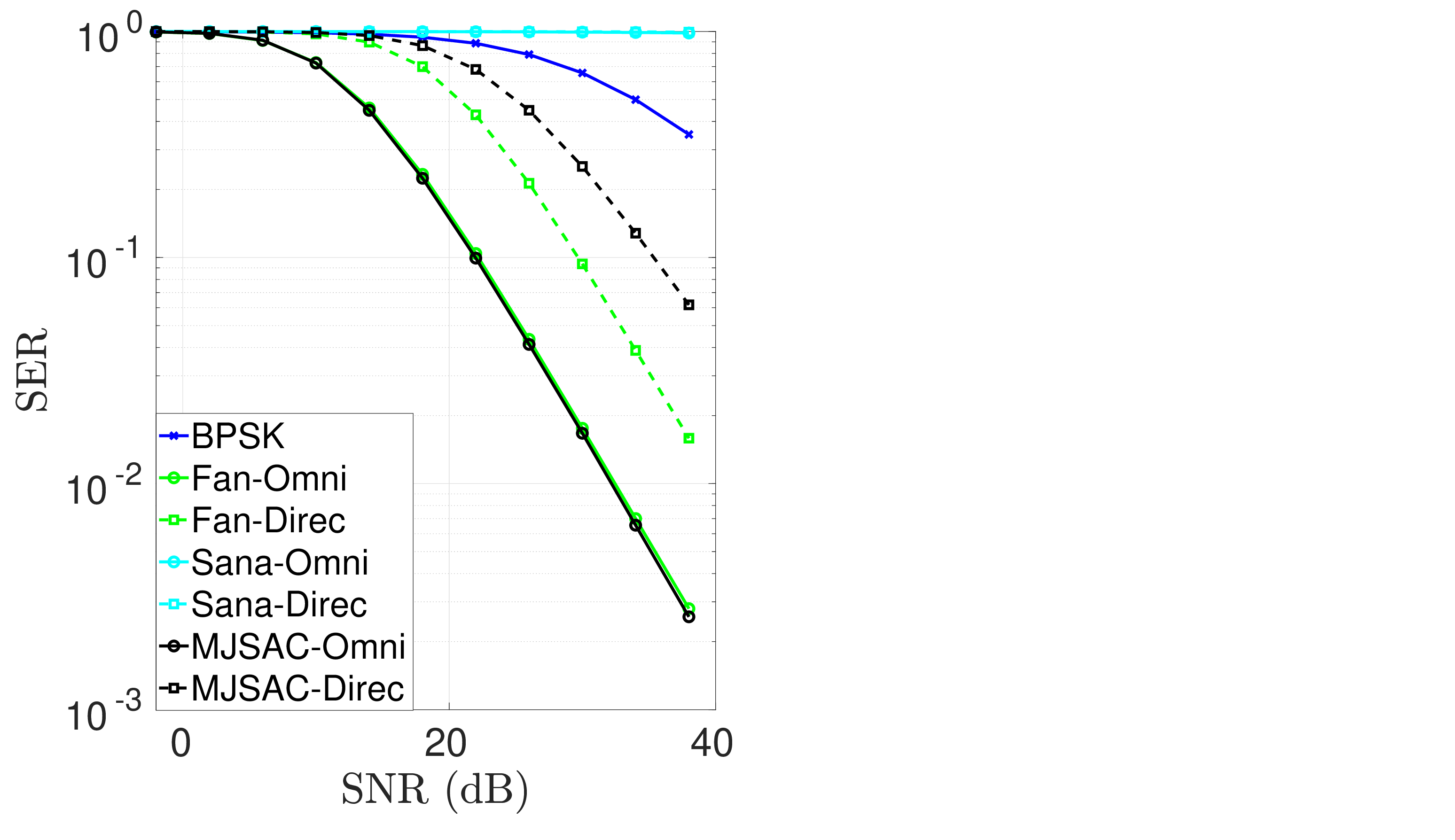}
\caption{Number of transmit antenna is 12.}
\label{fig:no_antennae12}
\end{subfigure}
\begin{subfigure}[]{.33\textwidth}
\centering
\includegraphics[width=1.8\linewidth,keepaspectratio]{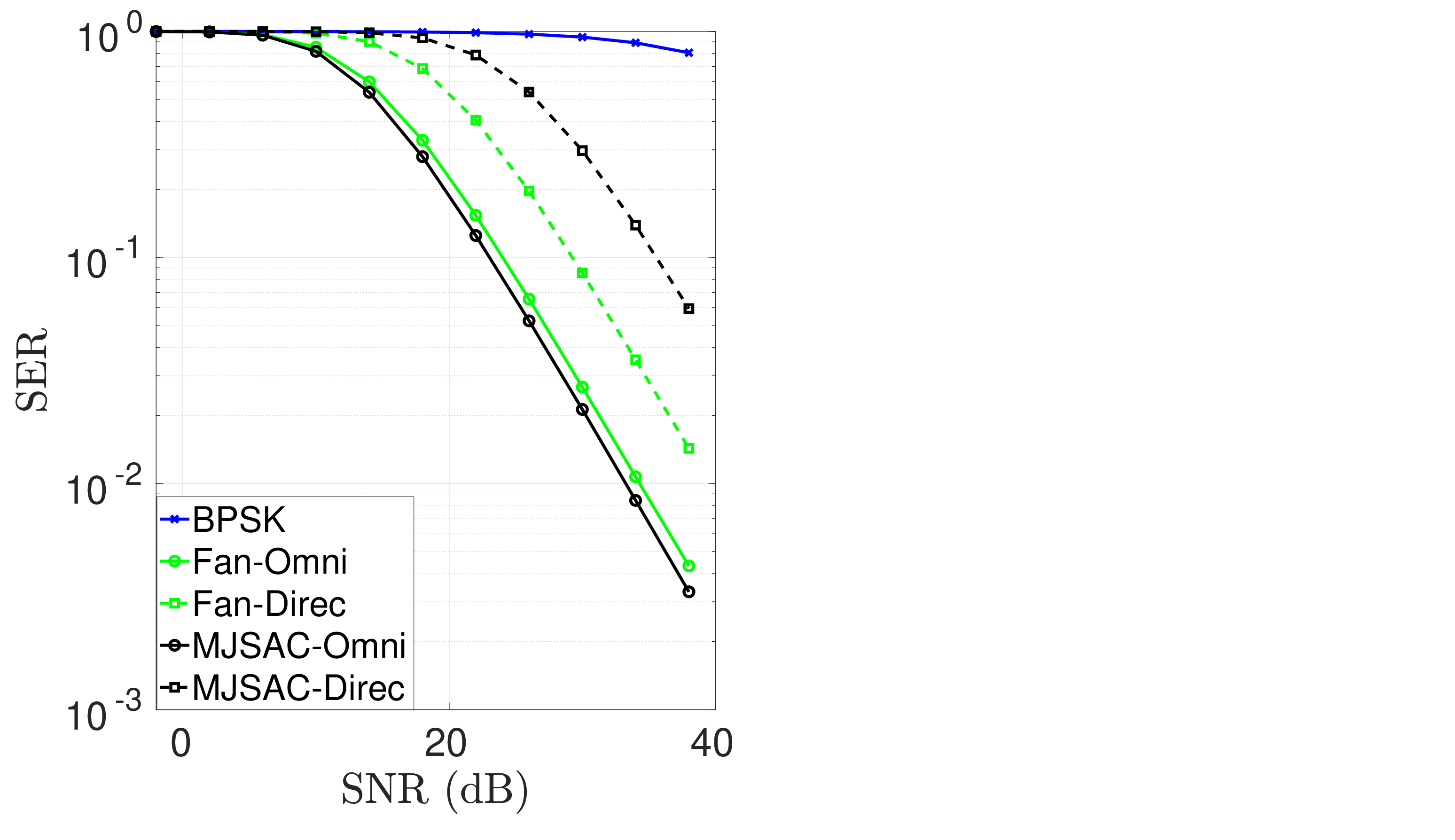}
\caption{Number of transmit antenna is 16.}
\label{fig:no_antennae16}
\end{subfigure}
\caption{Communication performance in terms of the symbol error rate.}
\label{fig:perf_BER_SER_bpsk_mcom}
\end{figure*}

\begin{figure}[!b]
\begin{subfigure}[]{0.45\textwidth}
\centering
\includegraphics[width=\linewidth,keepaspectratio]{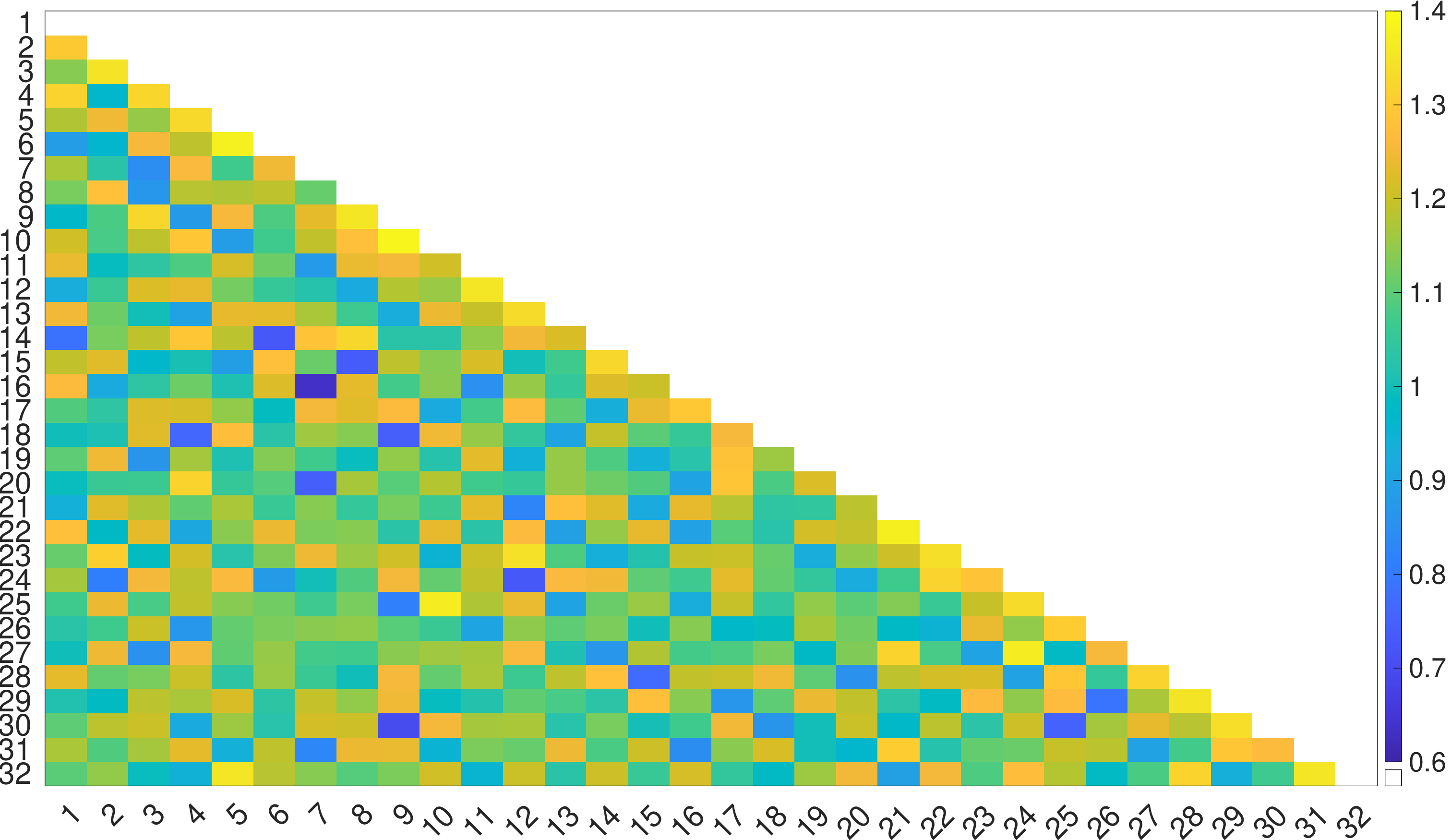}
\caption{\footnotesize Frobenius norm of differences among covariance matrices ($N=8$).}
\label{fig:Heat_map_ant8}
\end{subfigure}

\begin{subfigure}[]{.45\textwidth}
\centering
\includegraphics[width=\linewidth,keepaspectratio]{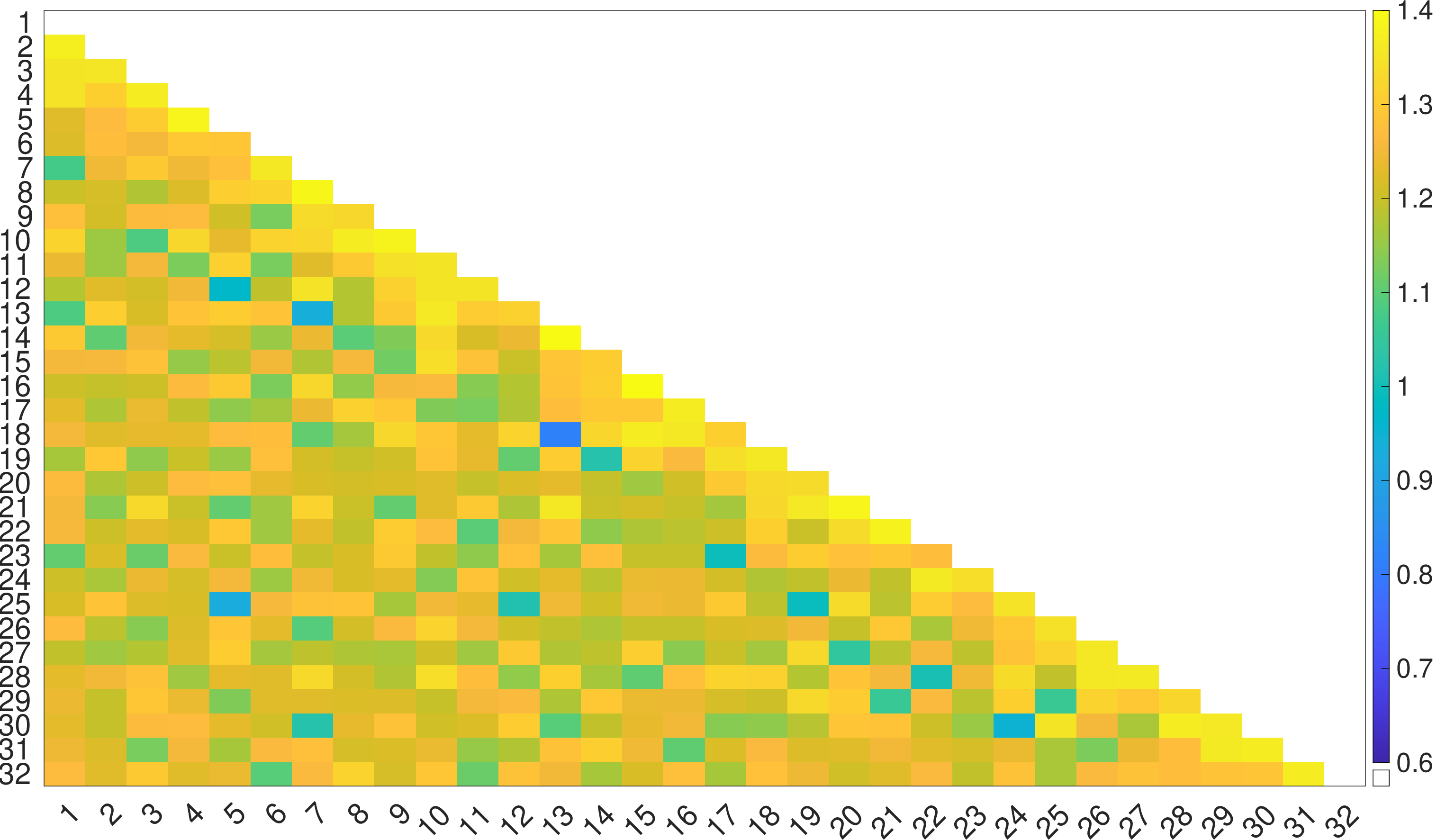}
\caption{\footnotesize Frobenius norm of differences among covariance matrices ($N=16$).}
\label{fig:Heat_map_ant16}
\end{subfigure}

\caption{Increase in the size of the covariance matrices leads to greater separation among them.}
\label{fig:heatmap}

\end{figure}

\color{black}
\section{Numerical Results}
\label{Sec_Numerical_result}
In this section, we will assess the performance of our proposed MJSAC and compare it with the existing techniques. For the simulation, we consider a scenario where a base station equipped with a uniform linear array of antennas transmits waveforms to achieve a desired beam pattern. The inter-element spacing between any two adjacent antennas is half of the wavelength. The transmitted waveforms also carry broadcast information symbols for several users located at different locations. Each user receives the transmitted signal using only one antenna and knows its angular location from the base station beforehand. For a fair comparison, the total transmit power for all techniques is set to unity. We compare the performance for two desired beam patterns: omnidirectional (Omni), where equal power is transmitted in all directions, and directional (Direc), where equal power is transmitted in three lobes: $\{-70^\circ, -50^\circ\} \cup \{-10^\circ, 10^\circ\} \cup \{50^\circ, 70^\circ\}$. We used an SDP-based technique \cite{stoica_probing_mimo_radar,Sajid_LiporTSP} to generate the covariance matrix, $\mathbf{R}$, corresponding to this beampattern. Figure \ref{fig:Beam_pattern} illustrates various beampatterns obtained for different numbers of transmit antennas. It is evident that as the number of transmit antennas increases, the desired beam pattern becomes better matched. 

In our first simulation, we conduct a comparative analysis of MJSAC's performance against binary phase shift keying (BPSK), and methodologies outlined in~\cite{Sana_2023} and~\cite{Liu_Fomni}. 
In terms of symbols, such a system can transmit $2^N$ symbols, each carrying $N$ bits. Therefore, for a fair comparison with BPSK, we design $2^N$ covariance matrices each carry a symbol or $N$ bits of data. The techniques presented in~\cite{Sana_2023} and~\cite{Liu_Fomni} are represented as Sana-Omni/Direc and Fan-Omni/Direc, respectively. To solve the optimization problem presented in \eqref{Eq:relax_McCormick}, we utilize the CVX toolbox \cite{grant2008cvx}, configuring $\alpha$ and $\beta$ as 1, as each element within $\mathbf{Z}_k$ is constrained to a value less than 1.  We examine three distinct scenarios, with $N$ set to 8, 12, and 16, respectively. Here, $L$ is defined as $2^N$. The communication frame's length, denoted as $M$, is standardized at 24, while $\bm{\phi}$ is implemented as a Hadamard matrix. Our simulation assumes the user's location at $\theta=60^\circ$.

The corresponding SER performance of all three schemes is shown in 
Fig.~\ref{fig:perf_BER_SER_bpsk_mcom}, It can be seen that our proposed technique MJSAC-Omni/Direc outperforms BPSK and Sana-Omni/Direc across all scenarios \footnote{We excluded Sana-Omni/Direc for $N=16$ due to its significantly poor performance.}. Furthermore, as $N$ increases, the performance gap between these techniques and MJSAC widens. The underperformance of Sana-Omni/Direc can be attributed to its failure to maximize the distance between the generated covariance matrices. Instead, it merely considers a convex combination of a predetermined set of covariance matrices. Fan-Omni surpasses MJSAC-Omni for $N=8$, however, for $N=12$ and $N=16$ MJSAC-Omni outperforms Fan-Omni. With an increase in $N$, the number of decision variables in~\eqref{Eq:relax_McCormick} rises, facilitating a significant distance between the symbols. However, increasing $N$ raises the complexity of the optimization problem.~\eqref{Eq:relax_McCormick}. Moreover, Fan-Omni/Direc requires the computation of the transmitted signal matrix for each transmission and also assumes that the transmitter possesses complete knowledge of $\mathbf{h}$. In contrast, MJSAC does not require the information of $\mathbf{h}$, and all the necessary symbols can be generated offline. The radar beam patterns obtained by MJSAC-Direc for different antenna sizes are shown in Fig.~\ref{fig:Beam_pattern}.

Fig.~\ref{fig:heatmap} illustrates the heat map depicting the Frobenius norm of the differences among 32 symbols $\left(S_i\right)$, obtained by MJSAC-Omni for $N=8$ and $N=16$. It is evident that there are many dark blue and green boxes (representing lower values of Frobenius norm) in Fig.~\ref{fig:Heat_map_ant8} as compared to Fig.~\ref{fig:Heat_map_ant16}.  It directly affects the communication performance as the receiver fails to detect the corresponding symbols correctly. This validates the poor performance of MJSAC-Omni when $N=8$ as compared to $N=16$ (refer Fig.~\ref{fig:no_antennae8} and Fig.~\ref{fig:no_antennae16}).

The worst-case complexity of an SDP~\cite{Pólik2010} is $ \mathcal{O}\left(I_1\left(p\sum_{i=1}^{N_{sd}}{d_{i}}^3 + p^2\sum_{i=1}^{N_{sd}}{d_{i}}^2 +  p^3\right)\right)$,
%
%
where $I_1$ is the iteration complexity, $p$ is the number of equality constraints, $d_i$ is the dimension of the $i^{\text{th}}$ semidefinite cone (SDC), and $N_{\text{sd}}$ is the number of SDC constraint. The computational complexity of MJSAC-Omni/Direc is $ \mathcal{O}\left( 4N^6 + 2N^7 + 8N^3\right)$. The computational complexity of Fan-Omni and Fan-Direc is $\sim$$\mathcal{O}\left(I_2NM^2)\right)$ and $\sim$$\mathcal{O}\left(I_2(NM^2+N^2M+N^3)\right)$ where $I_2$ is the total number of transmission, and $M$ is the length of the communication frame. MJSAC-Omni/Direc is computationally more expensive than Fan-Omni/Direc; however, its complexity is independent of $I_2$ and $M$. Furthermore, the computation of MJSAC-Omni/Direc is done offline, enhancing its scalability, whereas, in Fan-Omni/Direc, the computations are conducted online.

\section{Conclusion}

In this paper, we present a novel waveform design technique named MJSAC, specifically tailored for joint sensing and communication systems. Each symbol within the constellation creates a uniform predetermined beam pattern, making these waveforms suitable for both radar and communication applications. MJSAC solves an optimization problem to generate a set of covariance matrices. This is achieved by maximizing the distances (Frobenius norm) among these matrices while adhering to a set of constraints that ensure a consistent beam pattern. 
Notably, MJSAC is highly scalable as it does not consider channel information during symbol generation, allowing for offline computation. Extensive simulations confirm the superior performance of MJSAC across various scenarios compared to existing techniques.

\bibliography{IEEEabrv,ref}
\bibliographystyle{IEEEtran}

\end{document}